\documentclass[aps,twocolumn,amsmath,amssymb,preprintnumbers]{revtex4}
\usepackage{amsmath} \usepackage{amsfonts} \usepackage{amssymb}
\usepackage{bbm}
\usepackage{epsfig}
\usepackage{graphics}
\usepackage{graphicx}
\newcommand{\be}{\begin{equation}}
\newcommand{\ee}{\end{equation}}
\newcommand{\ba}{\begin{eqnarray}}
\newcommand{\ea}{\end{eqnarray}}
\newcommand{\nn}{\nonumber}

\newcommand{\mn}{_{\mu\nu}}

\begin{document}

\title[ ]{Universe without expansion}

\author{C. Wetterich}
\affiliation{Institut  f\"ur Theoretische Physik\\
Universit\"at Heidelberg\\
Philosophenweg 16, D-69120 Heidelberg}

\begin{abstract}
We discuss a cosmological model where the universe shrinks rather than expands during the radiation and matter dominated periods. Instead, the Planck mass and all particle masses grow exponentially, with the size of atoms shrinking correspondingly. Only dimensionless ratios as the distance between galaxies divided by the atom radius are observable. Then the cosmological increase of this ratio can also be attributed to shrinking atoms. We present a simple model where the masses of particles arise from a scalar ``cosmon'' field, similar to the Higgs scalar. The potential of the cosmon is responsible for inflation and the present dark energy. Our model is compatible with all present observations. While the value of the cosmon field increases, the curvature scalar is almost constant during all cosmological epochs. Cosmology has no big bang singularity. There exist other, equivalent choices of field variables for which the universe shows the usual expansion or is static during the radiation or matter dominated 
epochs. For those ``field coordinates`` the big bang is singular. Thus the big bang singularity turns out to be related to a singular choice of field coordinates.
\end{abstract}

\maketitle

After the discovery of general relativity, Einstein and others have tried to find static solutions of cosmology. This attempt has been abandoned after Hubble's observation of a systematic redshift proportional to the distance of a galaxy. This redshift has been taken as a clear indication for the expansion of distances with cosmic time. There is, however, a loophole in the argument. Imagine that masses of electrons and protons were smaller at the time of emission of radiation from a galaxy than they are today. Then the frequencies of characteristic atomic lines are also smaller than the ones observed on earth. This effect could replace the redshift due to expanding distances. In this note we demonstrate that such a scenario is perfectly viable. We construct a simple model for which cosmological distances shrink or remain constant. Only the Planck mass and the particle masses increase simultaneously with time \cite{CW1}. In contrast to earlier attempts in this direction \cite{Na1,Na2,Na3,Jain:2012cw} our 
model is compatible with all present cosmological observations, describing correctly nucleosynthesis or the emission of the cosmic microwave background. 

Our model predicts dynamical dark energy or quintessence \cite{CW3,RP,CW2,SL1,SL2,SL3,SL4,SL5} for late cosmology, while very early cosmology is characterized by an epoch of inflation. Inbetween one finds the usual radiation and matter dominated epochs. For all four periods the absolute value of the Hubble parameter $H$ remains almost constant, given by an intrinsic mass scale $\mu$. While the sign of $H$ is positive for inflation, it turns negative for radiation and matter domination. Nevertheless, we recover all standard predictions of cosmology. Since particle masses grow proportional to the Planck mass all observational bounds on the time variation of fundamental constants and apparent violations of the equivalence principle are obeyed. Despite similar predictions for observations our approach is not a simple reformulation of Einstein's general relativity. The cosmon field responsible for the change of masses plays a dynamical role, and its properties can be tested by observations of a dynamical dark 
energy 
or by the primordial density fluctuations from the inflationary period. 

The cosmology of our model has no big bang singularity. The field equations admit a solution which can be extended to infinite negative time $t\to-\infty$. In this limit the effective Planck mass and the scale factor approach zero. Invariants formed from the curvature tensor remain finite. 

Our model should be interpreted as a new complementary picture of cosmology, not as opposing the more standard picture of an expanding universe. The different pictures are equivalent, describing the same physics. This can be seen by a redefinition of the metric, which leads to the ``Einstein frame'' with constant Planck mass and particle masses and an expanding universe \cite{Di,CW1,Dam,Fla,Cat,Des}. In the Einstein frame the big bang has a singularity, however. The possibility of different choices of fields describing the same reality may be called ``field relativity'', in analogy to general relativity for the choice of different coordinate systems. Field relativity underlies the finding that strikingly different pictures, as an expanding or a shrinking universe, can describe the same reality. 

While the general setting of simultaneously varying Planck and particle masses, as well as the Weyl scaling to the equivalent Einstein frame, can be found in ref. \cite{CW1}, our simple picture of the Universe shrinking during radiation and matter domination is new and related to our specific model. Another  uncommon property of this model is the almost constant curvature scalar for all epochs. Furthermore, an important feature is the simplicity of our model covering both inflation and present dark energy, dominated by the same simple quadratic potential for the scalar cosmon field. No extremely small dimensionless parameter is introduced for the description of dark energy, the present value of the dark energy density in units of the Planck mass being tiny as a result of the large age of the Universe. 
Finally, the identification of the big bang singularity as a matter of the choice of field coordinates sheds new light on this old problem. 

\medskip\noindent
{\em Field equations.} The cosmological field equations can be derived by variation of the effective action $\Gamma$ which includes already all effects of quantum fluctuations. Our main points can be demonstrated for a simple form of the effective action for a scalar field $\chi$ - the cosmon - coupled to gravity,
\ba\label{1}
\Gamma=\int d^4x\sqrt{g}\left\{-\frac12\chi^2R+\frac12K(\chi)\partial^\mu\chi\partial_\mu\chi+\mu^2\chi^2 
\right\}.
\ea
The coefficient of the curvature scalar defines a dynamical Planck mass given by $\chi$, and we assume that all particle masses (except for neutrinos) are also proportional to $\chi$. For increasing $\chi$ the effective strength of gravity decreases $\sim\chi^{-2}$. The cosmon potential $V=\mu^2\chi^2$ will dominate the energy density both for the early inflationary epoch and for the present dark energy dominated epoch. The kinetic term leads to a stable theory for $K>-6~(K=-6$ is the ``conformal point''). Our choice of $K+6$ interpolates between a large constant $4/\tilde \alpha^2$ for $\chi^2\ll m^2$ and a small constant $4/\alpha^2$ for $\chi^2\gg m^2$. The detailed form of this interpolation does not matter.  To be specific, we take 
\be\label{2}
K(\chi)+6=\frac{4}{\tilde \alpha^2}\frac{m^2}{m^2+\chi^2}+\frac{4}{\alpha^2}\frac{\chi^2}{m^2+\chi^2}.
\ee
Compatibility with observations in late cosmology (bounds an early dark energy) requires $\alpha\gtrsim 10$, while a realistic inflationary period in early cosmology can be realized for small $\tilde \alpha$, say $\tilde \alpha=10^{-3}$. Other forms of a crossover from large values of $K+6$ for $\chi\to 0$ (not necessarily constant) to small values for $\chi\to\infty$ are viable as well. 

The present value of $\chi$ can be associated with the reduced Planck mass $M=2.44\cdot 10^{27}$eV, while the present value of $V=\mu^2\chi^2$ accounts for the dark energy density, such that $\mu\approx 2\cdot 10^{-33}$eV. Our model differs from a Brans-Dicke theory \cite{BD} by three important ingredients: the presence of a potential $V=\mu^2\chi^2$, the $\chi$-dependence of $K$ and, most important, the scaling of particle masses with $\chi$ \cite{CW1}. Since $\mu$ only sets the scale and can be taken 
as unity, the only three free parameters of the scalar and gravity part of our model are $\alpha$,$\tilde{\alpha}$ and $m/\mu$. 

For a homogenous and isotropic universe (and for vanishing spatial curvature) the field scalar and gravitational equations read \cite{CW1}
\ba\label{3}
K(\ddot{\chi}&+&3H\dot{\chi})+\frac12\frac{\partial K}{\partial \chi}\dot{\chi}^2= -\chi (2\mu^2-R)+q_\chi,\\
R&=&12H^2+6\dot{H}\nn\\
&=&4\mu^2-(K+6)\frac{\dot{\chi}^2}{\chi^2}-6
\frac{\ddot{\chi}}{\chi}
-18H\frac{\dot{\chi}}{\chi}
-\frac{T^\mu_\mu}{\chi^2},\label{4}\\
3H^2&=&\mu^2+\frac{K}{2}\frac{\dot{\chi}^2}{\chi^2}-6H\frac{\dot{\chi}}{\chi}+\frac{T_{00}}{\chi^2}.\label{B1}
\ea
The constant terms $~\mu^2$ on the r.h.s. of eqs. \eqref{4}, \eqref{B1} correspond to the potential divided by the squared Planck mass, $\mu^2=V/\chi^2$. As usual, we denote the scale factor in the Robertson-Walter metric by $a(t)$ and $H=\partial_t\ln a$. The energy momentum tensor $T\mn$ as well as $q_\chi$ reflect the effects of matter and radiation.

\medskip\noindent
{\em De Sitter solutions.} For constant $K$ one finds solutions where the geometry is for all times $t$ a de-Sitter space with constant $H$, while the effective Planck mass increases exponentially
\be\label{5}
H=b\mu~,~\chi=\chi_0\exp(c\mu t).
\ee
For $b>0$ the universe expands exponentially, for $b<0$ it shrinks exponentially. Insertion into eqs. \eqref{3}, \eqref{4}, \eqref{B1} yields algebraic equations for $b$ and $c$. The different cosmological epochs (inflation, radiation-, matter- and dark energy-domination) are characterized by different values of $b$ and $c$, always of order unity. Thus $\mu$ sets the characteristic time scale for the evolution in all epochs, including inflation and the vicinity of the big bang. Even though the scale factor and $\chi$ change exponentially, this change is very ``slow'', given by a characteristic time scale of the order of the present age of the Universe. Except for short transition periods numerical solutions are found to be well approximated by the solutions with constant $K$, where $K$ is evaluated for the appropriate value of $\chi$ \cite{CWNN}. 

In the absence of matter $(q_\chi=0,T^\mu_{\mu}=0)$ the combination of eqs. \eqref{3} and \eqref{4} yields two solutions, determined by
\ba\label{6}
(K+6)c^2_1&=&4~,~(K+6)c^2_2=\frac{4}{3K+16},\nn\\
3bc&=&\frac{2}{K+6}-2c^2.
\ea
The solution $c_1$ exists for all $K>-6$ with $bc=-2/(K+6)<0$. In this case one has
\be\label{7A}
c=\frac{2}{\sqrt{K+6}}~,~b=-\frac{1}{\sqrt{K+6}}=-\frac c2.
\ee
Furthermore, for $K>-16/3$ one has also the solution $c_2$ with 

\ba\label{9}
c=\frac{2}{\sqrt{(K+6)(3K+16)}}~,~b=\frac{K+4}{\sqrt{(K+6)(3K+16)}}.
\ea
Solutions with both $b$ and $c$ positive exist only for $K>-4$. For a scalar field dominated epoch radiation can also be neglected $(T_{00}=0$.) In this case only the solution $c_2$ \eqref{9} is consistent with eq. \eqref{B1}. Thus this solution is the one relevant for scalar field dominated cosmology. The solution $c_1$ \eqref{7A} is realized in the presence of radiation, see below.

\medskip\noindent
{\em Asymptotic initial cosmology.} We begin with scalar field dominated cosmology and assume $\tilde \alpha^2<2$ such that for $\chi\to 0$ the condition $K>-4$ is obeyed. Then scalar field dominated cosmology describes an exponentially expanding universe with exponentially increasing effective Planck mass $\chi$. As long as constant $K>-16/3$ remains a good approximation the solution \eqref{5}, \eqref{9} can perfectly describe the evolution of the universe for all times, including $t\to-\infty$. This solution is completely regular, no singularity is encountered. Indeed, it is easy to verify that the field equations have a solution which approach eqs. \eqref{5}, \eqref{9} for $t\to-\infty$, with $K+6=4/\tilde\alpha^2$. For the asymptotic past one has $\chi\to 0$ and the geometry is given by a de Sitter space with curvature tensor
\be\label{10A}
R_{\mu\nu\rho\sigma}=b^2\mu^2(g_{\mu\rho}g_{\nu\sigma}-g_{\mu\sigma}g_{\nu\rho}).
\ee
All invariants formed from the curvature tensor and its covariant derivatives are regular.

The ``big bang'' is free of any singularities. The central ingredient why the usual singularity is avoided arises from the behavior of the effective Planck mass $\chi$: it approaches zero as $t\to-\infty$. From the point of view of the field equations \eqref{3} this is in no way problematic, even though the effective strength of gravity, characterized by the effective Newton-constant $G(\chi)=1/(8\pi\chi^2)$, diverges for $t\to-\infty$. (A singularity free big bang has been observed in other contexts \cite{SW}, \cite{YP}).

\medskip\noindent
{\em Inflation.} We next show that the first stage of the evolution describes an inflationary universe. Let us take $\tilde \alpha \ll 1$. For the very early universe with $\chi\ll m$ one has $K+4=4/\tilde\alpha^2-2\gg 1$, such that $b\gg c$. In this case we can neglect $\ddot{\chi}$ as compared to $3H\dot{\chi}$ in eq. \eqref{3}. This property is called the ``slow roll approximation'' for inflation. We may continue the slow roll approximation to  larger values of $\chi$. As long as $\chi^2/m^2\ll \alpha^2/\tilde \alpha^2$ we can neglect in eq. \eqref{2} the term $\sim \alpha^{-2}$, such that the evolution equations read in the slow roll approximation
\ba\label{10}
H^2=\frac{\mu^2}{3}~,~
\dot{\chi}=\frac{\tilde \alpha^2\mu\chi(m^2+\chi^2)}{\sqrt{3}(m^2-3\tilde\alpha^2\chi^2)}.
\ea
The slow roll approximation breaks down once $\dot{\chi}/\chi$ is roughly of the same order as $H$. This happens for $\chi^2/m^2 \approx 1/(4\tilde\alpha^2)$ and we conclude that the inflationary slow roll phase ends once $\chi$ reaches a value of this order of magnitude. The amplitude of density fluctuations is governed by the ratio of the potential over the fourth power of the effective Planck mass, $\mu^2/\chi^2$. For large values of $m^2/(\tilde \alpha^2\mu^2)$ the density fluctuations can be very small, as required for a realistic cosmology. 

\medskip\noindent
{\em Radiation domination.} After the end of inflation entropy is created and the universe is heated. The subsequent radiation dominated period is realized for large $\chi$ where $K$ can be approximated by the constant ${4}/{\alpha^2} -6$. For radiation the trace of the energy momentum vanishes such that the field equations \eqref{3}, \eqref{4} are not  altered by the presence of radiation. However, eq. \eqref{B1} involves now the energy density of radiation $T_{00}=\rho_r$. Conservation of the energy momentum tensor implies $\rho_r\sim a^{-4}$ and therefore $T_{00}\sim \mu^4\exp (-4 b\mu t)$, or
\ba\label{B2}
\frac{T_{00}}{\chi^2}=\bar\rho_r\mu^2\exp \big\{-2(c+2b)\mu t\big\}.
\ea
For the shrinking universe according to the solution \eqref{7A} the energy density of radiation increases proportional to $\chi^2$, with $T_{00}/\chi^2=\bar\rho_r\mu^2$. Eq. \eqref{B1} is then obeyed for 
\be\label{B3}
\bar\rho_r=-3\frac{K+5}{K+6}.
\ee
A positive $\rho_r$ requires $K<-5$ or $\alpha^2>4$. The scenario of a shrinking radiation dominated universe with increasing effective Planck mass looks rather unfamiliar and intriguing. We will see below that this scenario predicts actually the same observations as the standard radiation dominated universe with expanding scale factor and constant Planck mass.

\medskip\noindent
{\em Matter domination.} The issue of matter is slightly more complicated. A realistic setting requires that the mass of the nucleon $m_n$ or the electron $m_e$ scale proportional to the growing Planck mass $\chi$. Otherwise the ratio $m_n/\chi$ would depend on time, violating the strict observational bounds. (Small deviations from this proportionality are allowed and could result  in an observable time variation of fundamental constants and apparent violation of the equivalence principle.) As a consequence of the proportionality of particle masses to $\chi$ one finds for massive particles an additional ``force" in eq. \eqref{3}, adding a term $q_\chi=-(\rho-3p)/\chi$ on the right hand side \cite{CW1}. Also on the r.h.s of eqs. \eqref{4}, \eqref{B1} one has now to add terms $-T^\mu_\mu/\chi^2=(\rho-3p)/\chi^2$ and $T_{00}/\chi^2=\rho/\chi^2$, respectively. For a conserved particle number the density $n$ is diluted as $n\sim a^{-3}$. Thus the energy density of a pressureless gas will scale $\sim\chi a^{-3}$ 
and therefore become comparable to radiation at some time. After this matter-radiation equality we can essentially neglect radiation and follow the evolution in a matter dominated universe. 

For $\rho\sim\chi^2$ (and neglecting $p$) the additional terms on the r.h.s. of the field equations are constant (after dividing eq. \eqref{3} by $\chi$). Solutions of the type \eqref{5} are again possible, now with $\rho=\bar \rho_m\mu^4\exp \big\{(-3b+c)\mu t\big\}$,
\ba\label{B5}
\frac{\rho}{\chi^2}=\bar\rho_m\mu^2~,~3b+c=0.
\ea
With the addition of the corresponding terms on the r.h.s eqs. \eqref{3}, \eqref{4}, \eqref{B1} are all obeyed for constant $K$ and
\ba\label{B6}
c=\sqrt{\frac{2}{K+6}},~\qquad~b=-\frac{1}{3}\sqrt{\frac{2}{K+6}}=-\frac13c,
\ea
with 
\be\label{B7}
\bar\rho_m=-\frac{2(3K+14)}{3(K+6)}.
\ee
This solution exists for $K<-14/3$ or $\alpha^2>3$. 

At this point cosmology is described by a sequence of three de Sitter geometries, all with exponentially increasing $\chi$. For the first scalar dominated inflationary period the universe is expanding, while it shrinks for the subsequent radiation and matter dominated epochs. The Hubble parameter $H=b\mu$ remains always of the same order of magnitude, changing sign, however, after the end of inflation. 

\medskip\noindent
{\em Dark energy domination.} In the present epoch we live in a transition from the matter dominated era to an epoch dominated by dark energy.  This can be triggered by neutrinos becoming non-relativistic, provided the neutrino mass grows faster than $\chi$, e.g. $m_\nu\sim\chi^{(2\tilde\gamma+1)}$. Such a scenario of growing neutrino quintessence \cite{ABW,CWNEU} resembles closely the $\Lambda$CDM cosmology for redshift $z\lesssim 5$. For constant $\tilde\gamma$ the future is again given by a de Sitter solution with 

\ba\label{B7a}
b&=&\frac{(2\tilde\gamma-1)c}{3}~,~c^2=\left(\frac{K+6}{2}+\frac{4\tilde\gamma(1-\tilde\gamma)}{3}\right)^{-1},\nn\\
\rho_\nu&=&\bar\rho_\nu\mu^2\chi^2~,~\bar\rho_\nu=\frac{8(1+\tilde\gamma)-6(K+6)}{8\tilde\gamma(1+\tilde\gamma)+3(K+6)},
\ea
such that the universe expands for $\tilde \gamma>1/2$.

Growing neutrino quintessence is rather economical in our context. The same potential $V=\mu^2\chi^2$ describes inflation and late dark energy and no new parameters need to be introduced in the scalar-gravity sector. Additional parameters involve only the $\chi$-dependence of the neutrino masses, for which $\tilde \gamma$ may actually depend on $\chi$. For example, the model of ref. \cite{CWNEU} introduces only two additional parameters - the present average neutrino mass and the scale $\chi_t$ where neutrino masses grow large. Together with $\alpha, \tilde{\alpha}$ and $m/\mu$ this is a rather minimal five-parameter set for an overall description from inflation until now. (Additional ''particle physics parameters`` determine $\Omega_m$ and $\Omega_b$, the relation between temperature and $\rho_r$ or the heating of the Universe after inflation.) 

As an alternative to growing neutrino masses the kinetial $K$ could be modified such that for the present value of $\chi$ it gets large again \cite{HW}. For example, $K$ could be periodic in $\chi$. For large $K$ the scaling solution with matter \eqref{B6}, \eqref{B7} is no longer possible and the universe may return to a scalar field dominated cosmology according to the solution \eqref{7A}.

\medskip\noindent
{\em Einstein frame.} The units in which geometric distances are measured can be changed by a field dependent redefinition of the metric (conformal transformation)
\be\label{18A}
g_{\mu\nu}=\left(\frac{M^2}{\chi^2}\right)^\eta g'_{\mu\nu}.
\ee
This is a change of ''field coordinates'', not to be confounded with a usual general coordinate transformation. Field transformations change the form of the effective action and the field equations. Nevertheless, the choice of field variables does not matter for physical observables. A particularly useful choice is the Weyl scaling to the Einstein frame, $\eta=1$. Using for the scalar field the variable
\be\label{12A}
\varphi=\frac{2M}{\alpha}\ln \left(\frac\chi\mu\right)
\ee
the action \eqref{1} reads in the Einstein frame (omitting primes on $\sqrt{g}, R$ and $\partial^\mu$)
\ba\label{A2}
&&\Gamma=\int d^4x\sqrt{g}\left\{-\frac{M^2}{2}R\right.\nn\\
&&\quad \left.+\frac{k^2}{2}\partial^\mu\varphi\partial_\mu\varphi+M^4\exp \left(-\frac{\alpha\varphi}{M}\right)\right\},\nn\\
&&k^2=\frac{\alpha^2(K+6)}{4}.
\ea
The Planck mass is now given by a constant $M$. Particle masses that scale $\sim \chi$ in the ``Jordan frame'' (using $g\mn$) are constant in the ``Einstein frame'' (using $g'\mn$) \cite{CW1}.  The discussion of observations is typically most easily done in the Einstein frame since one has no longer to pay attention to a varying gravitational constant and varying particle masses. We can use the Einstein frame in order to establish that our model is compatible with present observations. 

For large $\chi$ where $k^2\approx 1$ the effective action \eqref{A2} describes a standard model for quintessence with an exponential potential. One recovers the known scaling solutions for the radiation $(n=4)$ and matter $(n=3)$ dominated epochs, with a constant fraction of early dark energy $\Omega_h={n}/{\alpha^2}$ \cite{CW3,CW2}. One may verify that the de Sitter solutions \eqref{7A}, \eqref{B3} and \eqref{B6}, \eqref{B7} are in one to one correspondence with these scaling solutions. The predictions from nucleosynthesis or the emission of the background radiation are only slightly influenced by the presence of early dark energy. This can be used for establishing bounds on $\alpha$. Dynamical dark energy  takes the standard form of growing neutrino quintessence \cite{ABW,CWNEU}.

The inflationary period occurs for small or negative $\varphi$. It is described by ``cosmon inflation'' \cite{CI} and we refer to this work for a quantitative discussion. In particular, one finds that the slow roll parameters obey at horizon crossing the relations $\epsilon=\eta=\tilde \alpha^2\chi^2/(2m^2)=1/(2N)$, with $N$ the number of $e$-foldings before the end of inflation. For the primordial density fluctuations this leads to a prediction \cite{CI} of the spectral index $n=0.97$ and the scalar to tensor ratio $r=0.13$. The determination of cosmological parameters by the Planck collaboration \cite{PL}, $n=0.96\pm 0.01$, is consistent with this prediction. The value $r=0.13$ may be considered borderline, requiring an analysis in the presence of the massive neutrinos and early dark energy of growing neutrino quintessence. (Modifications of the effective action for small $\chi$ can also realize a smaller value of $r$ \cite{CI}.)  The observed amplitude of density fluctuations measures the parameter 
combination $\tilde\alpha^2\mu^2/m^2\approx (2/3)\cdot 10^{-10}$, resulting in a second characteristic mass scale $\hat m=2m/\tilde \alpha\approx 5\cdot 10^{-28}$eV besides $\mu$. 

\medskip\noindent
{\em Absence of big bang singularity}. At this point it may be worthwhile to discuss the origin of the apparent singularity of the big bang in the Einstein frame for the metric. Approaching the big bang for $t'\to 0$ one finds that $H'$ diverges as $H'=(1+b/c)/t'=2/(\tilde \alpha^2 t')$, with a corresponding divergence of the curvature scalar $R'=48/(\tilde\alpha^4 t'^2)$. For the solutions discussed in this note we can associate the ``big bang'' with $\chi\to 0$. This happens for $t\to -\infty$ in the Jordan frame and for $t'\to 0$ in the Einstein frame. The curvature scalar formed from the metric $g\mn$ remains finite, cf. eq. \eqref{10A}, \eqref{4}, while it becomes singular for the metric $g'\mn$ at the time $t'\rightarrow 0$. The reason is simply that $R'$ is related to $R$ by a multiplicative factor $M^2/\chi^2$ which diverges for $\chi\to 0$. (The precise relation contains also additive terms involving derivatives of $\chi$.) We conclude that the usual ``big bang singularity'' is a ``coordinate effect'
' in the space of field variables 
originating from a singular field transformation. There exists a simple choice of fields where the big bang solution is regular for all time. 

Physical observables should not depend on the choice of fields used to describe them. One may call this property ``frame invariance''. If there exist ``physical singularities'' concerning observables they can be detected in all frames - typically indicating a shortcoming or incompleteness of the theoretical model. In contrast, a singularity that appears for one choice of fields but is absent for at least one other choice may be called a ``field singularity'', in analogy to a ``coordinate singularity'' that may appear in certain coordinate-systems. The cosmological solution  \eqref{5}, \eqref{9}, \eqref{10A} shows no singularity for arbitrary $t$. The singularity appearing for this solution in the Einstein frame is therefore a field singularity. This property does not hold, however, for all homogenous and isotropic solutions in the Jordan frame. There exists a class of solutions, neighboring the cosmological solution for late time, which cannot be continued to $t\to-\infty$. This feature is directly related 
to 
the stability of the cosmological solution \cite{CWNN}.

One may ask if physical singularities may be encountered in other physical observables as the big bang is approached, perhaps related to perturbations rather than to the ``background geometry''. For example, the propagation of gravity waves becomes singular at the big bang in the Einstein frame, and one may therefore suspect an associated physical singularity that also should be visible in the Jordan frame. Interestingly, the graviton remains no longer a propagating mode in the Jordan frame in the big-bang-limit $\chi\to 0$. Indeed, its kinetic term, proportional to the prefactor of $R$ in eq. \eqref{1}, vanishes for $\chi\to 0$. (This is analogous to the absence of propagating gluons in the confinement regime of QCD.) For our specific model the singularity of the graviton propagation in the Einstein frame is related to the absence of a propagating graviton in the Jordan frame. So far we have not found any physical observable which becomes singular for $\chi\to 0$. We conjecture that the big bang is actually 
free of 
physical singularities in our model. 

Having discussed two different frames for the description of the same universe it is clear that other choices are possible as well. Different choices of $\eta$ in eq. \eqref{18A} yield different expansion histories. For $\eta=1/2$ the geometry is static flat Minkowski space for the radiation dominated period, while a static Minkowski geometry can be realized during matter domination if one takes $\eta=1/3$. There exists a choice of $\eta$ for which the universe is static at present, shrinking in the past and expanding in the future. For general $\eta\neq 0$ the big bang is singular, such that a regular cosmology singles out the frame with action \eqref{1}. An exception is $\eta =1-2/\tilde\alpha^2$ for which the geometry of the big bang is flat Minkowski space. More details, as well as different models in the spirit of this note, can be found in ref. \cite{CWNN}.

In conclusion, we have constructed a ``variable gravity universe'' whose main characteristic is a time variation of the Planck mass or associated gravitational constant. The masses of atoms or electrons vary proportional to the Planck mass. This can replace the expansion of the universe. A simple model leads to a cosmology with a sequence of inflation, radiation domination, matter domination, dark energy domination which is consistent with present observations. The big bang appears to be free of singularities.

\bibliography{universe_without_expansion}

\end{document}